\documentstyle[aps,preprint,psfig]{revtex}
\begin {document}
\title {\bf Fermi Statistics of Weakly Excited Granular Materials \\ 
in a Vibrating Bed II: One Dimensional Experiment}
\author {Nathan C. Blanchard, 
Paul V. Quinn \thanks{E-mail address: pvq2@lehigh.edu}, 
Daniel Ou-Yang \thanks{E-mail address: hdo0@lehigh.edu}, \\
Joseph Both \thanks{E-mail address: jabc@lehigh.edu},  and 
Daniel C. Hong \thanks{E-mail address: dh09@lehigh.edu}}
\address{
Department of Physics, Lewis Laboratory,
Lehigh University, Bethlehem, Pennsylvania 18015}
\date {\today}
\maketitle
\begin {abstract}
A one dimensional experiment in granular dynamics is carried out to test the 
thermodynamic theory of weakly excited granular systems [Hayakawa and Hong, 
Phys. Rev. Lett. {\bf 78}, 2764(1997)] where granular particles are treated 
as spinless Fermions.  The density profile is measured and then fit to the 
Fermi distribution function, from which
the global temperature of the system, $T$, 
is determined.  Then the center of mass, $< z(T) >$, and its fluctuations, 
$<(\Delta z(T))^2>$, are measured and plotted as functions of $T$.  The 
Fermi function fits the density profile fairly well, with the value of $T$ 
being fairly close to the predicted value. The 
scaling behavior of $< z(T) >$ and $<(\Delta z(T))^2>$ is 
in excellent agreement with the theory.
\end{abstract}
\vskip 0.2 true cm
\noindent PACS numbers: 81.05Rm, 05
\vskip 2.0 true cm
\noindent{\bf I. The Experimental Setup} 

This paper is a sequel to the preceeding paper [1], where
the thermodynamic theory of Hayakawa and
Hong[HH] [2] was tested 
with extensive Molecular Dynamics simulations.  The purpose of this paper
is to test the theory of HH experimentally for  
a one dimensional vibrating granular system.  A one dimensional system is
perculiar in the sense that randomness associated with collisions is
suppressed in contrast with what happens in
higher dimensions.  Nevertheless, 
the Fermi statistics arising from hard core repulsion
still apply to this simple one dimensional system.
As in the preceeding paper,
we determine the configurational statistics of a one dimensional
vibrating granular system by properly taking the ensemble average of
the steady state.  Then we measure the following quantities and
compare them to those predicted by HH.  First, we measure the
density profile and determine the {\it dimensionless} Fermi
temperature $T=T_f/mgD$($T_f$ is the Fermi temperature) 
by fitting the profile to the
Fermi function.  Second, we compare the measured Fermi temperature $T$
to those predicted by the theory (Eq.(6a) of the preceeding paper.)
Third, we measure the center of mass and the fluctuations of the
vibrating bed and test the scaling predictions (Eqs. (5c) and (7) of the 
preceeding paper.)
Since the detailed theoretical aspect can
be found in the preceeding paper, we only discuss here the experimental
aspect and the analysis of the experimental data
as well as its comparison with the
theoretical predictions.

We now explain the experimental setup in some detail.  The one 
dimensional column of particles used in this experiment was 30 plastic 
beads.  The beads, having holes through their centers, were strung through a 
thin piece of copper wire.  The wire was stretched extremely tight between two 
horizontal rods clamped to a ring stand.  Each of the beads could freely move 
up and down the copper wire.  The beads were all close to the same size, with 
an average diameter of D=5.74 mm.  
Differences in the diameter of the beads were
not noticeable to the naked eye.  The bottom of the copper wire ran through
a small metal bar connected horizontally to a mechanical vibrator, very 
similar to an audio speaker.  One of the plastic beads was glued to the metal 
bar, acting as the bottom wall when the vibration was turned on.  A 
schematic diagram of 
the set up is shown in Fig.1.

The mechanical vibrator was then
connected to a function generator, which allowed 
one to control the type, amplitude, and frequency of the mechanical 
vibration.  For this experiment, the only type of vibration used was a cosine 
wave, $A\cos(\omega t)$, where $A$ is the amplitude and $\omega = 2 \pi f$ is 
the angular frequency.  The frequency of the vibration and the voltage which 
set the amplitude were both controlled with the function generator.   
The relationship between the voltage and the vibration amplitude may be 
nonlinear and complex.  Hence, instead of deriving the mathematical formula,
for a given applied voltage,
we measured directly the corresponding amplitude of the vibration by
putting one bead on the vibrating plate and measured its maximum height.
We find that the
voltage with a range of 0.00 V to 9.50 V corresponds 
to an amplitude of 0.000 cm to 0.584 cm.  In this experment, the frequency 
$f$
was kept at $f$=40Hz while the voltage was varied by 2 volts from 1.00 V to 
9.00 V.  The corresponding amplitudes, A,  and the vibration strength
($\Gamma=A\omega^2/g$) with g the gravitational acceleration are listed in 
Table I.  This set of
control parameters satisfies the
criterion of the weakly excited granular system used to test the 
theory of HH(Eq.(3) of the preceeding paper), namely:
$$ R=\mu_o D/\Gamma A >>1 \eqno (1)$$
with $\mu_o$ being the initial number of layers.
To analyze the system, pictures were taken with a digital camera while the 
beads were subjected to vibration.  The camera was placed on a stand in front 
of the beads, and a large convex lens was placed in between to focus the image 
properly.  This is schematically
illustrated in Fig.2.  A set of 32 pictures was taken for 
each value of $\Gamma$ used in this experiment, including a picture of the 
beads at rest in the begining and the end of each run.  Then, the images were 
scanned into the computer so that the center of each bead in every picture 
could be obtained.  
Once the individual bead coordinates were known, they were fed into 
various FORTRAN programs and used to calculate the density profile, the center 
of mass, and the fluctuations of the center of mass for each voltage setting.

To find the density profile, equally sized boxes with a height equal to 
the diameter of a bead were constructed from the
grain at the bottom, so that each box contained, one bead in the ground 
state.  
Note that the reference point in the density measurement is not the vibrating
plate but the particle at the bottom, which {\it is} fluidized.  Before
proceeding, we want to clarify two points regarding the density measurement:
First,
on the density plot presented in this paper, the density could exceed one
for certain data points.  This is because box sizes were assumed to all have
a length and width equal to the {\it average} bead radius.  In reality,
however, this was not true, leading to a density value that can be greater 
than 1.  Second, since the bottom layer is
the reference point, one may expect that the
density of the first data point should remain the same, but one may notice
that the values of the density of the bottom data changes slightly.  This is
due to the erosion of the balls that
occurred due to the repeated motion on the copper wire.
This erosion of the center
hole caused the configuration of the bottom few balls to be more dense
as more trials were completed.

For this 
experiment, 50 boxes were chosen because the motion of the column of beads 
never reached that height, even for the largest voltage.  Therefore, 
every bead was included in the calculations of all the density 
profiles.  Since this is a one dimensional experiment, each bead was 
considered to be a circle instead of a sphere.  Hence, the maximum possible 
area of a bead in a box, $a_o$, is $\pi r^2$ where $r=<D>/2=\sum_iD_i/2N$ 
is the average
radius of a bead with $D_i$ the diameter of each bead and N the total number 
of beads, i.e, N=30. In the ground state, when 
the system is at rest, each box up to the Fermi surface
contains one bead with 
an area of $a_o$.  However, when the beads are subjected to vibration, 
the system expands, and the time averaged position of each grain rises.  The
area of the grains in each box is computed from the pictures,
which is denoted as $a$. 
Then, the density 
of each box, having a value between zero and one, was computed using 
$$\rho = \frac{a}{a_o}.$$
This process was done for all 50 
boxes and in all 32 pictures for each voltage used.  
An average density 
for each box was then found by summing up the 32 densities per box and 
dividing by 32.  The density profile for a particular voltage was obtained 
by plotting the average density of each box against the box number.
The center of the first box is chosen as the origin.
The profile is then fit
with the Fermi function as discussed in the previous section 
and a global temperature, $T$, is obtained.  The fitting of the density 
profile was done with a non-linear least squares program as well as the eye.  

FORTRAN programs were used to obtain the center of mass and its 
fluctuations.  The center of mass is computed with the following formula:
$$z = \frac{\Sigma_{i} z_i m_i}{\Sigma_{i} m_i},$$
where $z_i$ is the vertical position in centimeters of the $i$th bead and 
$m_i$ is the mass in grams of the $i$th bead.  These summations were carried 
out over all 30 beads.  This formula was used to get the center of mass for 
each picture, and then the average was taken over all 32 pictures to get 
the average center of mass, $<z>$.  The average fluctuations of the center 
of mass are computed using the following standard formula for the deviations:
$$< (\Delta z)^2 > = \frac{\Sigma_{j}(z(j) - < z >)^2}{32},$$
where $z(j)$ is the center of mass of an individual picture, and $< z >$ 
is the average center of mass found previously for a particular setting of 
the voltage.  In this case, the summation is carried out over 32, the number 
of pictures taken per run.  Both the center of mass and the fluctuations were 
computed for each voltage setting.
\vskip 2.0 true cm

\noindent{\bf II. Data and Results}

{\bf Density Profile}:
This experiment was carried out for 30 particles using a cosine wave vibration 
for voltages of 1.00 V, 3.00 V, 5.00 V, 7.00 V, and 9.00 V at a constant 
frequency of $f$ = 40 Hz.  The change in $V$ is synonamous with a change in 
the vibration strength, $\Gamma$.  Density profiles, $\rho$, as a function of
vertical position z, for the different voltages were fit to the Fermi profile 
and can be seen in Fig.3a-e.  Note that
for an electron gas, the Fermi profile is given by:
$$\rho(\epsilon_i) = 1/(1+ exp(\epsilon_i - \bar\mu)/T_f) \eqno (2a)$$
where $\epsilon_i$ is the i-th energy level of the electrons, and 
$\bar\mu\equiv mg\mu(T) D$  is the
Fermi energy and $T_f$ is the Fermi temperature.  As discussed in [1],
for granular systems, $\mu(T)$ is independent of the temperature because
the density of the state is constant.  We denote this constant as $\mu_o
\equiv \mu(T=0)$, which is the initial number of layers.  Note that 
the dimensionless Fermi energy at T=0 is the initial number of layers, i.e.,
$\mu(T=0)\equiv\mu_o$.
For the configurational
statistics of granular materials in a vibrating bed, the
energy level is given by the average {\it position} of the particle.
Thus, the energy level is given by 
the gravitational energy, $\epsilon_i = mgDz_i\equiv mgD\bar
z$ where $z_i$
is the dimensionless vertical position of the i-th grain
measured in units of D.  Note that $\bar z$ runs
from 0 to 50
because the position of the first grain is taken as the origin.
The relationship between the {\it dimensionless} 
fitting Fermi temperature
$T$ in this paper and the
Fermi temperature $T_f$ in (2a) is given by $T=T_f/mgD$:
$$ \rho(\bar z) = 1/(1+exp([\epsilon_i(\overline{z})-\bar\mu)/T_f])\equiv
1/(1+exp((\bar z-\mu)/T) 
\eqno (2b) $$

All the density profiles plotted together can be 
seen in Fig.4. 
Note that the Fermi statistics are valid when the
the parameter R defined by (Eq.(3)) in the preceeding paper [1] is greater
than 1, namely: $R = \mu D/\Gamma A$.  
In our experiment, R changes from 131.07(voltage = 1.00 
V) to .9327(voltage = 9.00 V)(See Table II).

{\bf Fermi temperature fitting}:
When fitting the experimental data with the Fermi function, the following two 
adjustable parameters are used:  the Fermi energy, $\mu$, and the temperature, 
$T$.  The parameter $\mu$ shifts the location of the Femi energy 
horizontally, and the temperature, $T$, controls the curvature around 
the Fermi energy.  As discussed above, the Fermi energy is expected 
to remain constant regardless of the temperature $T$
(If the density of states is constant, i.e,
$\mu(T)= \mu_o$).  However, the 
experimental fitting shows that there is some slight variation in $\mu$.  In 
this experiment, $\mu$ varies from 29.6 to 32.0, a relative increase of 
about $\Delta \mu/\mu = 0.08$.  The scaling behavior of the center of mass and 
its fluctuations don't seem to be affected by the change in $\mu$. However, 
there is a large affect in the amplitude as
we have experienced in the Molecular Dynamics simulations [1].
The fitting values for $\mu$ and  
$T$ are listed in Table II along with the ratio $R$.  When the  
temperature, $T$, obtained by the Fermi fitting is compared to the 
to the theoretical prediction, $T_o$,(Eq.6a) of the preceeding paper, 
they seem to match fairly well.  For the sake of completeness, we present
this formula again:
$$ T_o = =\frac{1}{D}\frac{T_f}{mg} = \frac{1}{D}
\sqrt{\frac{6D(gH_o(\Gamma)/\omega^2)}
{\pi^2\alpha}} \eqno (3)$$
A best fit value was $\alpha=1.5$.  Translating this into a physical
picture, we notice that
the relationship between the actual
amount of expansion in the center of
mass in the one dimensional vibrating bed, $\Delta h$, and the
jump height, $\bar h=gH_o/\omega^2$, of a single ball 
fired gently(meaning the
relative velocity and relative acceleration between the ball
and the plate are both zero)
to the free surface by the vibrating plate is given by the relation:
$\Delta h=\bar h/\alpha$.
Note that the last point in Table II is off, for which $R \approx 0.93 <1$.
This is expected since 
the Fermi analysis is no longer valid at such a high voltage.  
Table III shows the values of both the predicted temperature, $T_o$, and
the fitting temperature $T$.  They agree fairly well except for the last 
one.    

{\bf The center of mass}:
The relative increase in the
center of mass of the one dimensional column is denoted as $<\Delta z(T)>$, 
which is
the difference in the actual position of the center of mass, $z(T)$, and that
of the ground state, i.e.,
$\Delta z(T)= z(T)-z(T=0)$ is given by(Eq.(5) of ref.[1]):
$$ < \Delta z(T)> = \frac{\pi^2 D}{6\mu}T^2 \eqno (4)$$
where $\mu$ is the dimensionless Fermi energy(the initial number of layers.)
Note the appearance of the factor D in (4).  It
is because,
in ref. [1], the temperature T has the dimension of length, while here
it is dimenionless.  
The center of mass is plotted in Fig.5 as a function of $T^2$ in a graph of 
five different values of $\Gamma$.  By using the solid line as a guide for the 
eye, one can see that the graph seems to confirm the scaling predictions of 
the Fermi statistics, $\Delta z \propto T^2$.  Notice that the last point 
deviates dratically from the best fit line.  Once again, this is because the 
conditions of the system violate the inequality (1)
and make the Fermi analysis invalid.  
Hence, this point was not included in obtaining the best fit line.  There is 
a discrepency in the proportionality constant C.  Theory 
predicts that one should get $ C = \pi^2D/6\mu \approx 0.031$ 
The experimental results 
yield, $C \approx .45$, a difference of about a factor 14
or more when compared 
to the theory.  It has been shown in a preceeding paper [1]
that this discrepancy is due to the extreme sensitivity of the center of mass, 
to the Fermi energy, $\mu$.  When the density of 
states is independent of the energy, the Fermi energy must remain constant.  
However, in this one dimensional experiment, it changes for each value of 
$\Gamma$.  This small change does not seem to change the Fermi fitting in 
Fig.3a-e, but it does effect the amplitude of the scaling relationships.

{\bf Fluctuations of the center of mass}:
The fluctuation of the center of mass, $<(\Delta z)^2>$, is given by(Eq.(7) of
ref.[1]):
$$ <(\Delta z)^2 >= \frac{\pi^2}{3}\frac{D^2}{\mu^2}T^3 \eqno (5)$$
The fluctuations of the center of mass were also measured and plotted as a 
function of $T^3$ in Fig.4.  Using the guide line, the graph once again seems 
to confirm the validity of the Fermi statistics which imply that $
<(\Delta z)^2> \propto T^3$.  The proportionality constant is way off from the 
theoretical value, 
$D^2\pi^2/3\mu^2 \approx 10^{-3}$.  
From the graph, 
the slope is approximately 0.15.  The last point was again
eliminated when obtaining 
the best fit line, just as it was for the center of mass for the reason
explained above.  
Considering that the amplitude of the center of 
mass is off by a factor of 14, 
one should expect that the fluctuations would be 
off by a factor of at least $14^2$.  
This is, again, due to the sensitivity of the Fermi 
integral to $\mu$, where small changes in $\mu$ are greatly magnified in the 
fluctuations, resulting in the factor of order $10^2$ 
difference.  We have found a similar trend in the MD simulations [1].
Another source of 
error is due to the fact that the fluctuations in the center of mass are quite 
large in the vibrating column, and all the particles, not just those near the 
surface, fluctuate in the continuum space of the experiment.  This is different
from the Fermi model which makes all particles below the Fermi surface, 
inactive.  So the positions of the grains, on average, may obey the Fermi 
distribution function well, but the magnitude of the fluctuations may not.  It 
is very suprising, though, that the $T^3$ prediction of the Fermi statistics 
still seems to hold.
\vskip 2.0 true cm 

\noindent {\bf III. Conclusion}

We now summarize
the main results of this experiment as follows.  {\it First}, the 
configurational statistics of grains in a one dimensional system subject to 
vibration seem to obey the Fermi statistics of spinless particles for weakly 
excited systems as was predicted in the paper by HH[2].  {\it
Second,} the temperature 
determined by fitting the Fermi profile is fairly close to the theoretical 
prediction.  
{\it Third,}
the scaling relations of the center of mass and its fluctuations obey the 
Fermi statistics, but there are discrepencies in the amplitudes or 
proportionality constants.  Note that another source of error 
in this experiment results from friction.
The beads were confined to vibration in a single column by the 
copper that ran through their centers.  After many trials, filings from the 
inside of the plastic beads were found at the base of the setup.  The motion 
of the beads on the wire eroded away some of the hole in the center of each 
bead.  This friction could be responsible for some of the error in the 
results.  
Nevertheless, we find that the essence of the Fermi statistics survives in
this one dimensional experiment.
\vskip 2.0 true cm

\noindent {\bf Acknowledgement}

This research was supported by NSF as a part of Research
Experiences for Undergraduate Students program at Lehigh University.

\newpage
\noindent {\bf References}
\vskip 0.2 true cm
\noindent [1] P. Quinn and D. C. Hong, preceeding paper and
references therein.
\vskip 0.2 true cm
\noindent [2] H. Hayakawa and D. C. Hong, Phys. Rev. Lett. {\bf 78}, 2764 
(1997).
\newpage

\noindent {\bf Figure Captions}
\vskip 1.0 true cm
\noindent Fig.1. The schematic picture of the
experimental setup used to vibrate a column of beads at 
a controlled amplitude and frequency.  
\vskip 0.3 true cm
\noindent Fig.2. The schematic picture of the
experimental setup used to take pictures of the vibrating 
column of beads with the digital camera.  
\vskip 0.3 true cm
\noindent Fig.3. Density profiles of the one dimensional vibrating beads
and the fitting to the Fermi function.  The voltages used in each run are:
(a)1.00 V, (b)3.00 V, 
(c)5.00 V, (d)7.00 V, and (e)9.00 V.  A plot of all the profiles at once is 
given in (f).
\vskip 0.3 true cm
\noindent Fig.4 The center of mass, $<z(T)>$, is plotted as a function of 
$T^2$. The straight line is a guide to the eye.  
\vskip 0.2 true cm
\noindent Fig.5. Fluctuations in the center of mass, $<(\Delta z(T)^2>$, are 
plotted as a function of $T^3$. The straight line is a guide to the eye.  
In both Figs. 4 and 5, the 
last point was not included in the estimation of the best fit line, because
it violates the criterion given by Eq.(2).

\newpage
\noindent {\bf Table Captions}
\vskip 1.0 true cm
\noindent Table I. The values of the vibrational amplitude A, and the
vibration strength $\Gamma$ for 
each corresponding voltage V. 
\vskip 1.0 true cm
\noindent Table II.  A set of parameters used in the experiment for 
various voltages.  
Here, $\mu$ is the dimensionless Fermi
energy, $T$ is the dimensionless
Fermi temperature, and $R=\mu D/\Gamma A $ is the
parameter that determines the validity of the Fermi statistics.(Eq.1)
\vskip 1.0 true cm
\noindent Table III. The comparison between
the theretical Fermi temperature, $T_{o}$ and the
experimentally measured Fermi temperature, $T$, for different vibrations at 
voltage V.

\newpage
\thispagestyle{empty}
\centerline{\hbox{
\psfig{figure=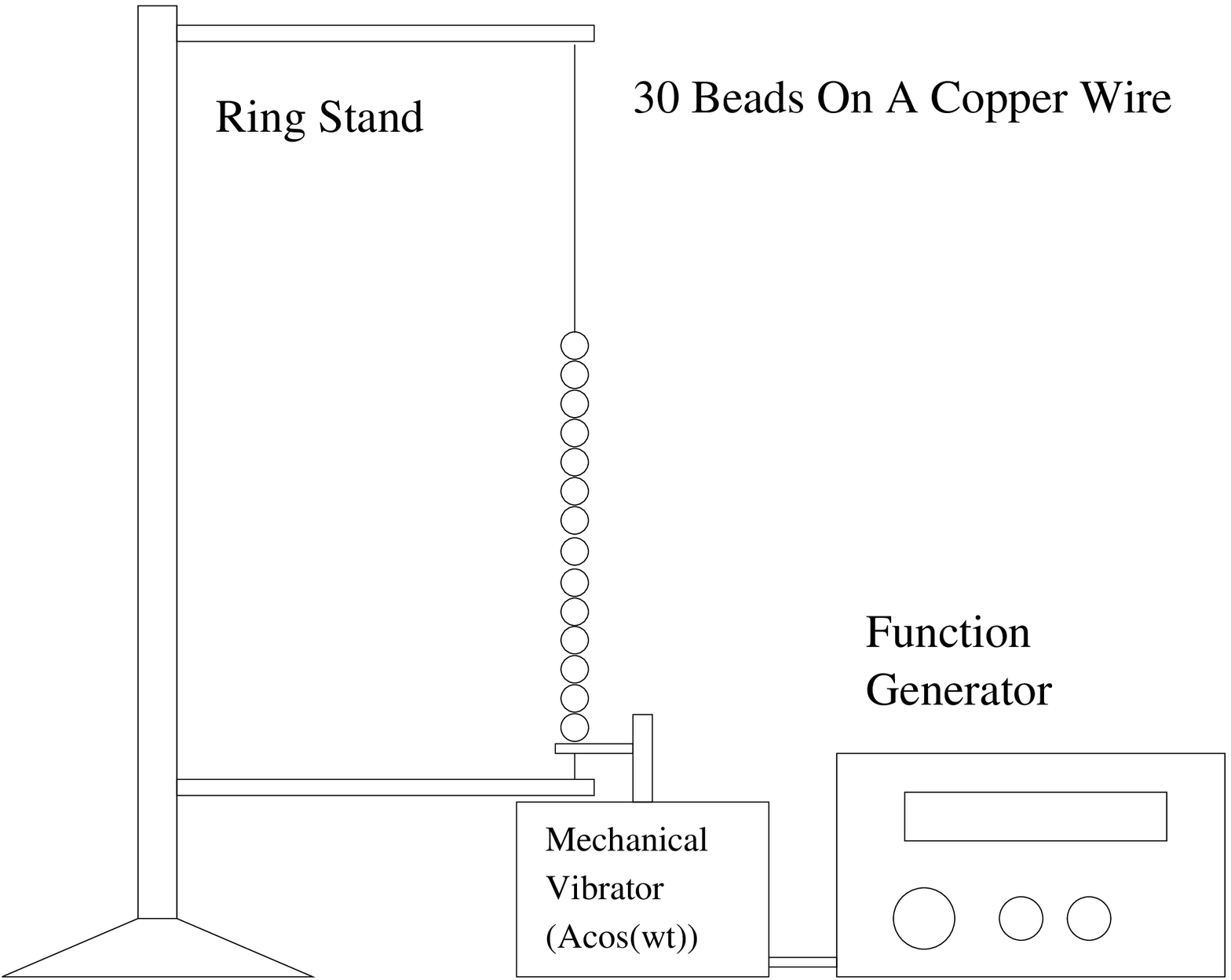}
}}

\newpage
\thispagestyle{empty}
\centerline{\hbox{
\psfig{figure=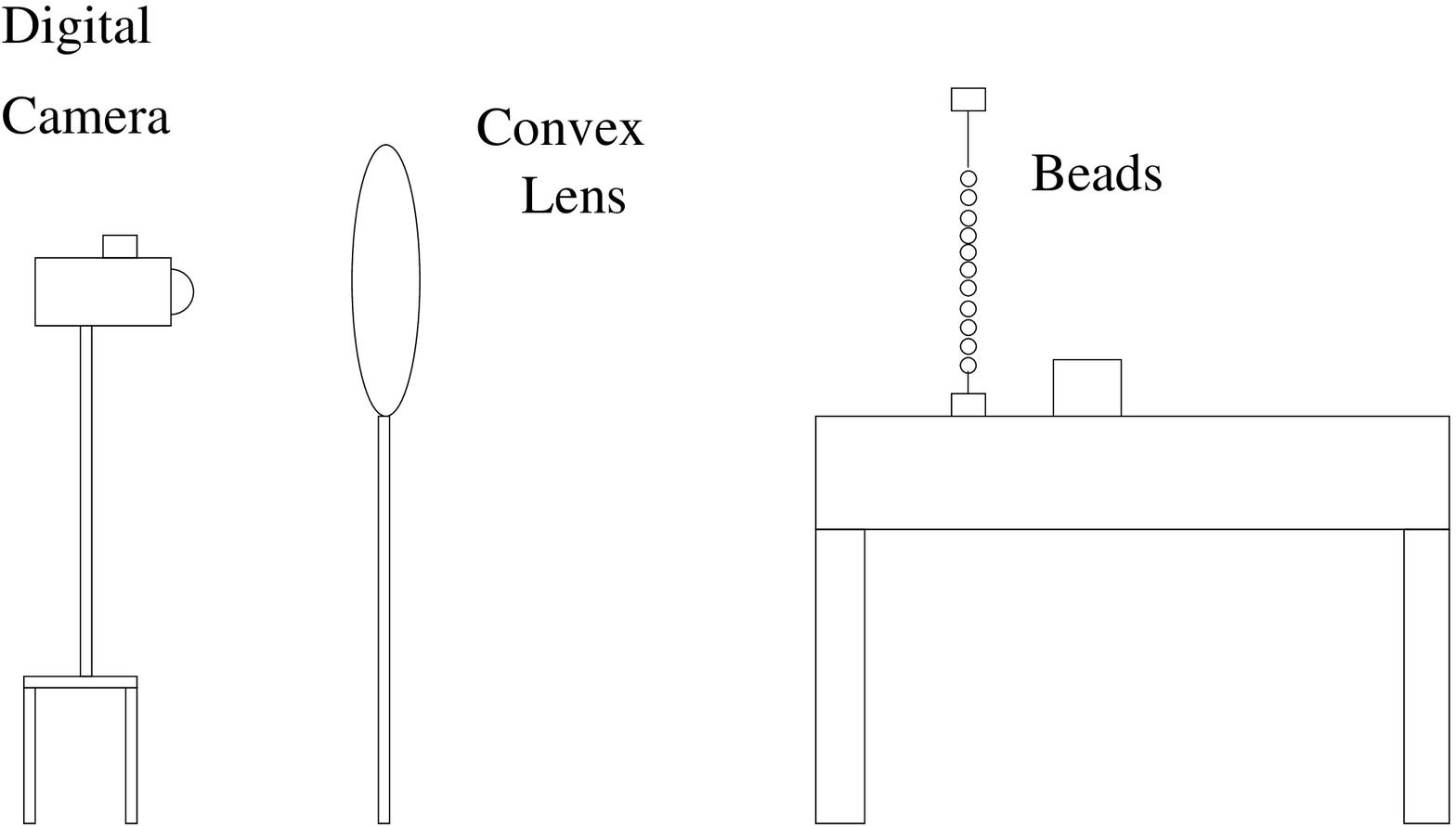}
}}

\newpage
\thispagestyle{empty}
\centerline{\hbox{
\psfig{figure=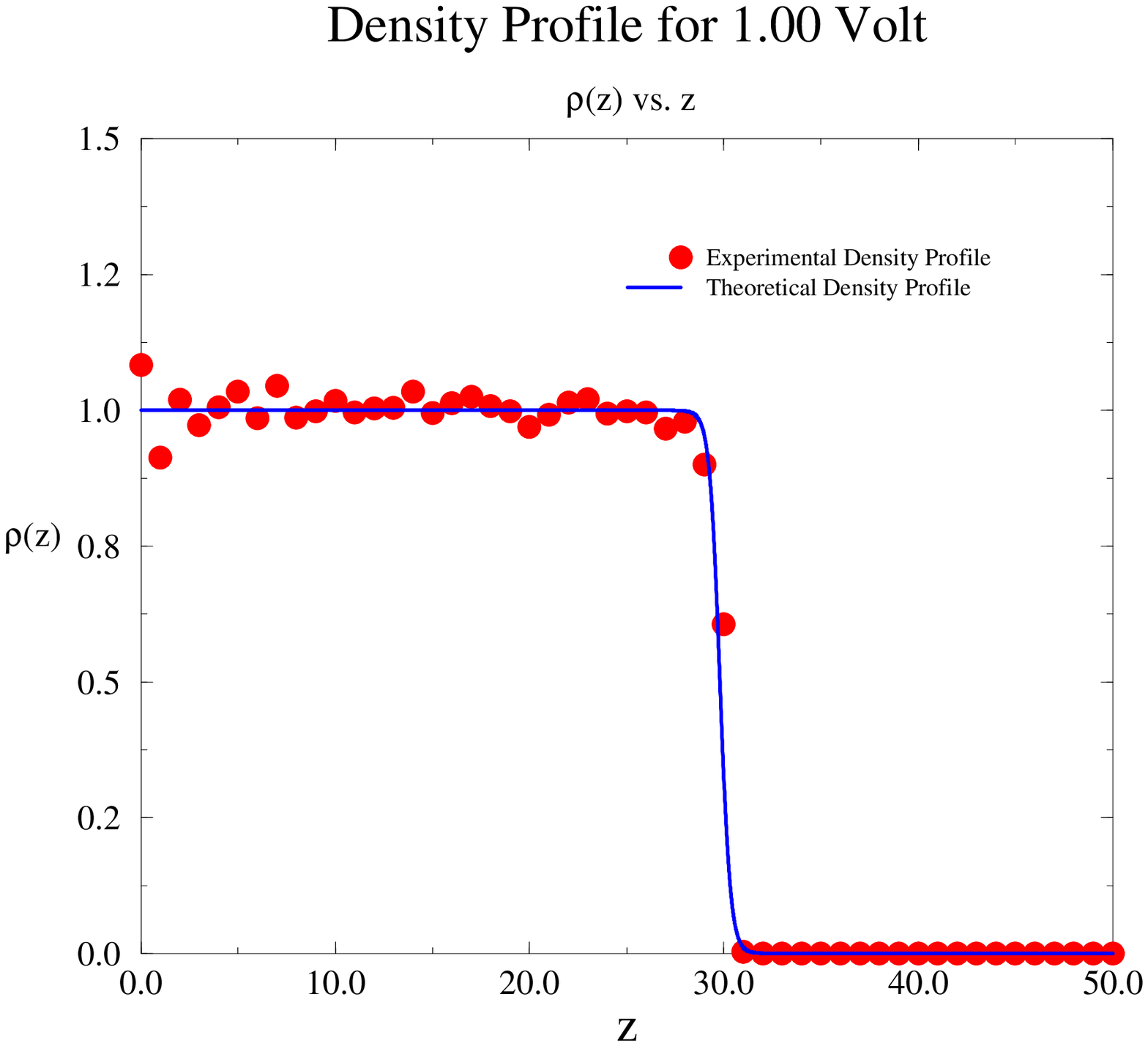}
}}

\newpage
\thispagestyle{empty}
\centerline{\hbox{
\psfig{figure=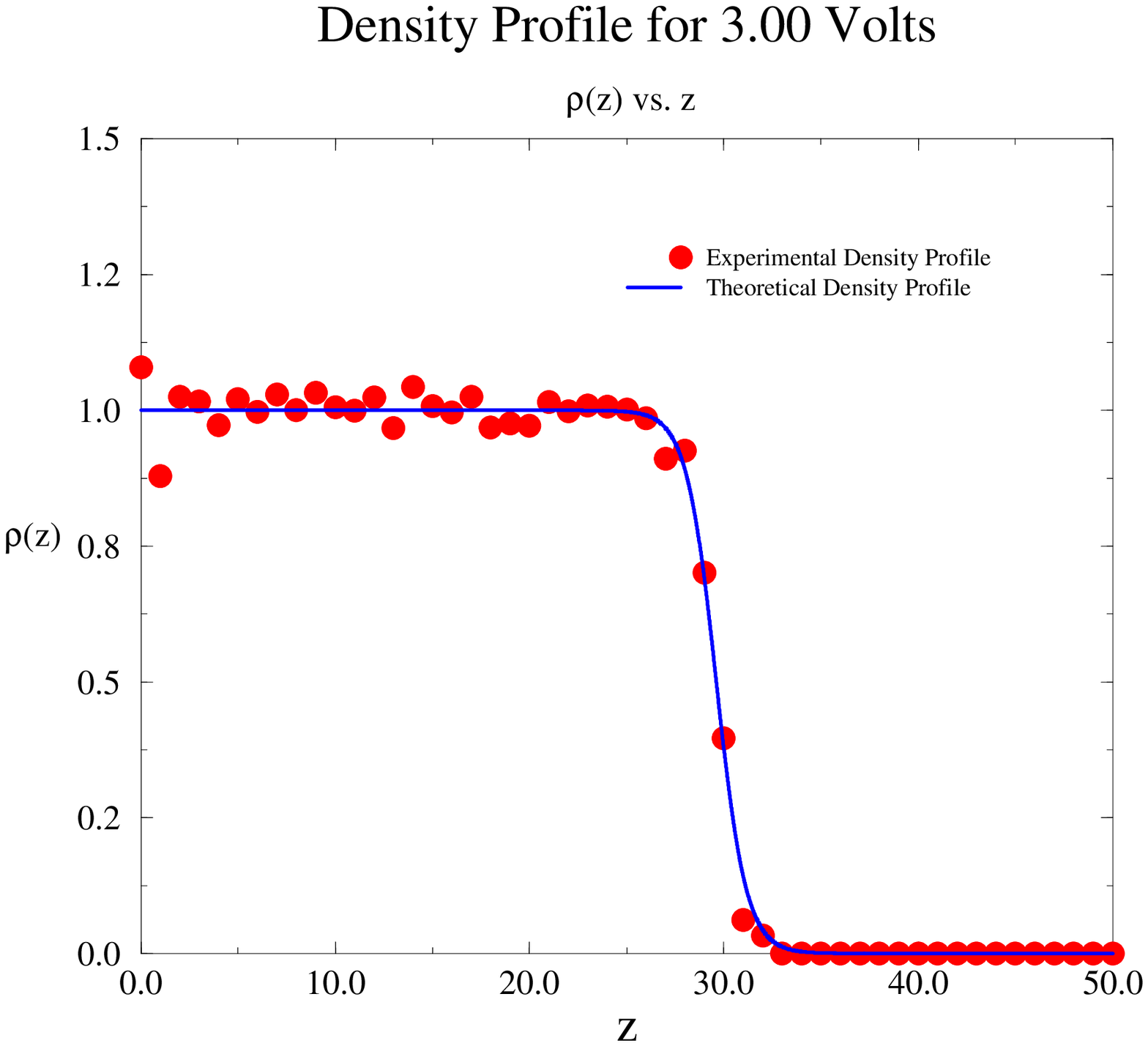}
}}

\newpage
\thispagestyle{empty}
\centerline{\hbox{
\psfig{figure=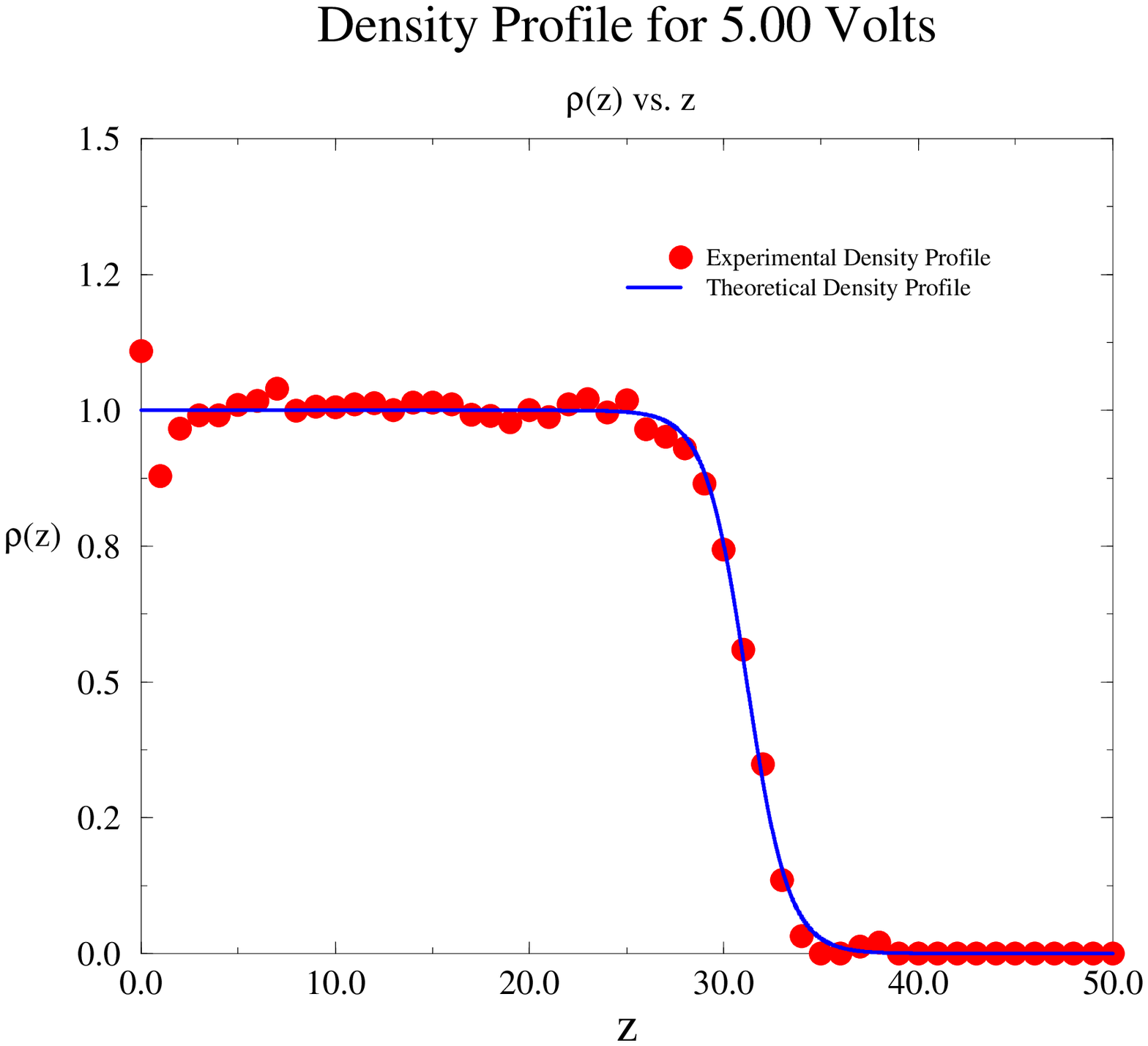}
}}

\newpage
\thispagestyle{empty}
\centerline{\hbox{
\psfig{figure=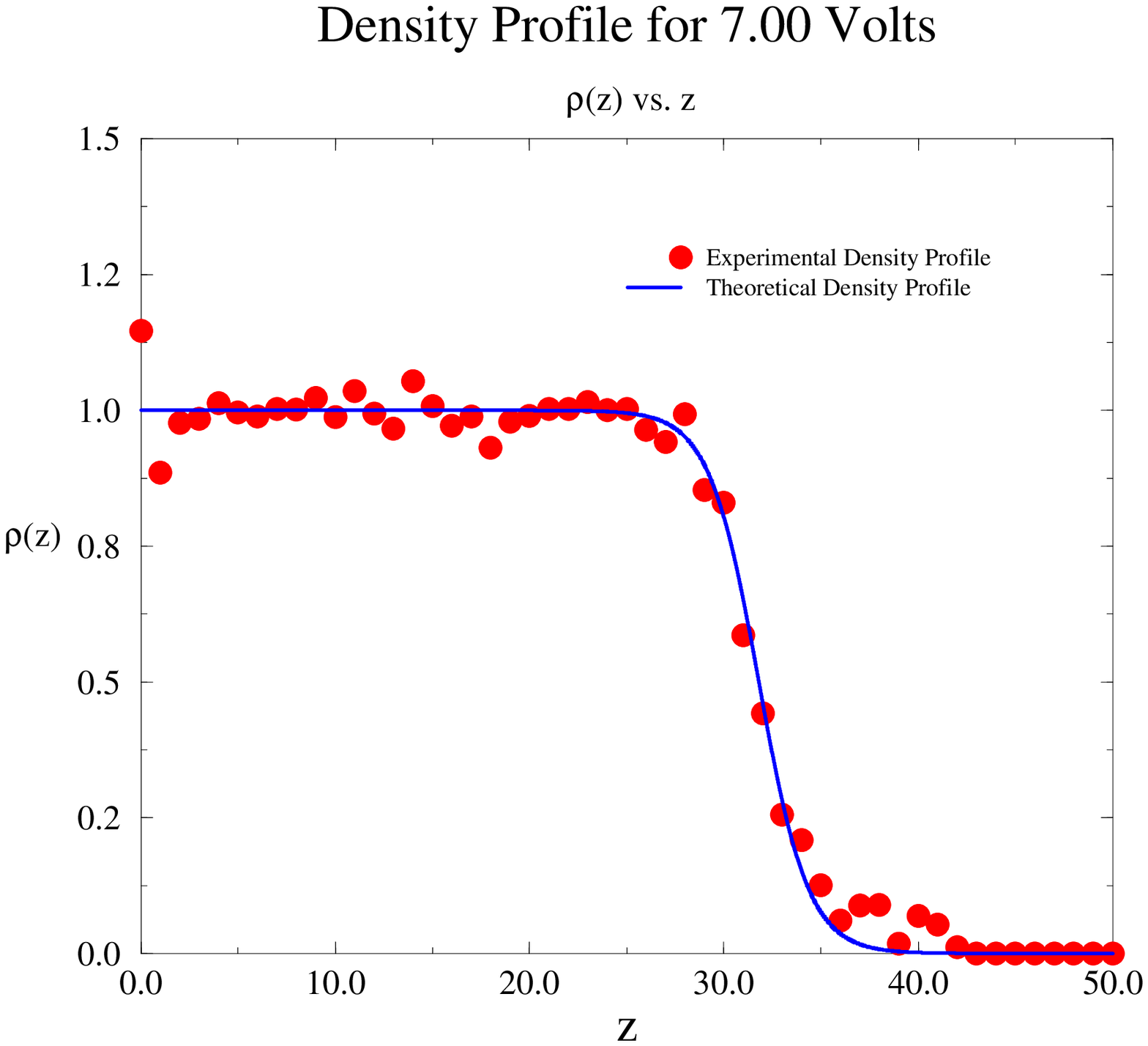}
}}

\newpage
\thispagestyle{empty}
\centerline{\hbox{
\psfig{figure=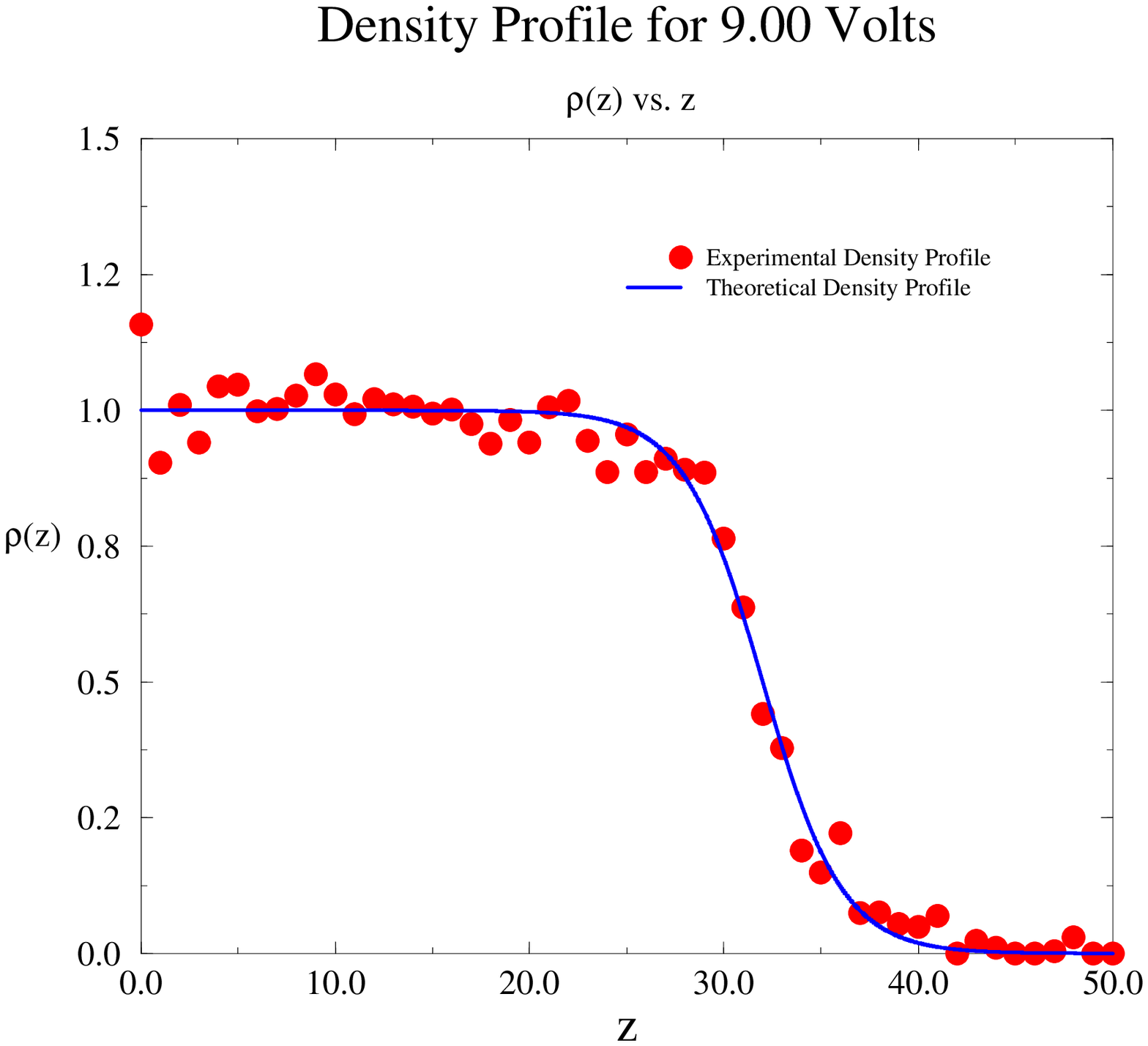}
}}

\newpage
\thispagestyle{empty}
\centerline{\hbox{
\psfig{figure=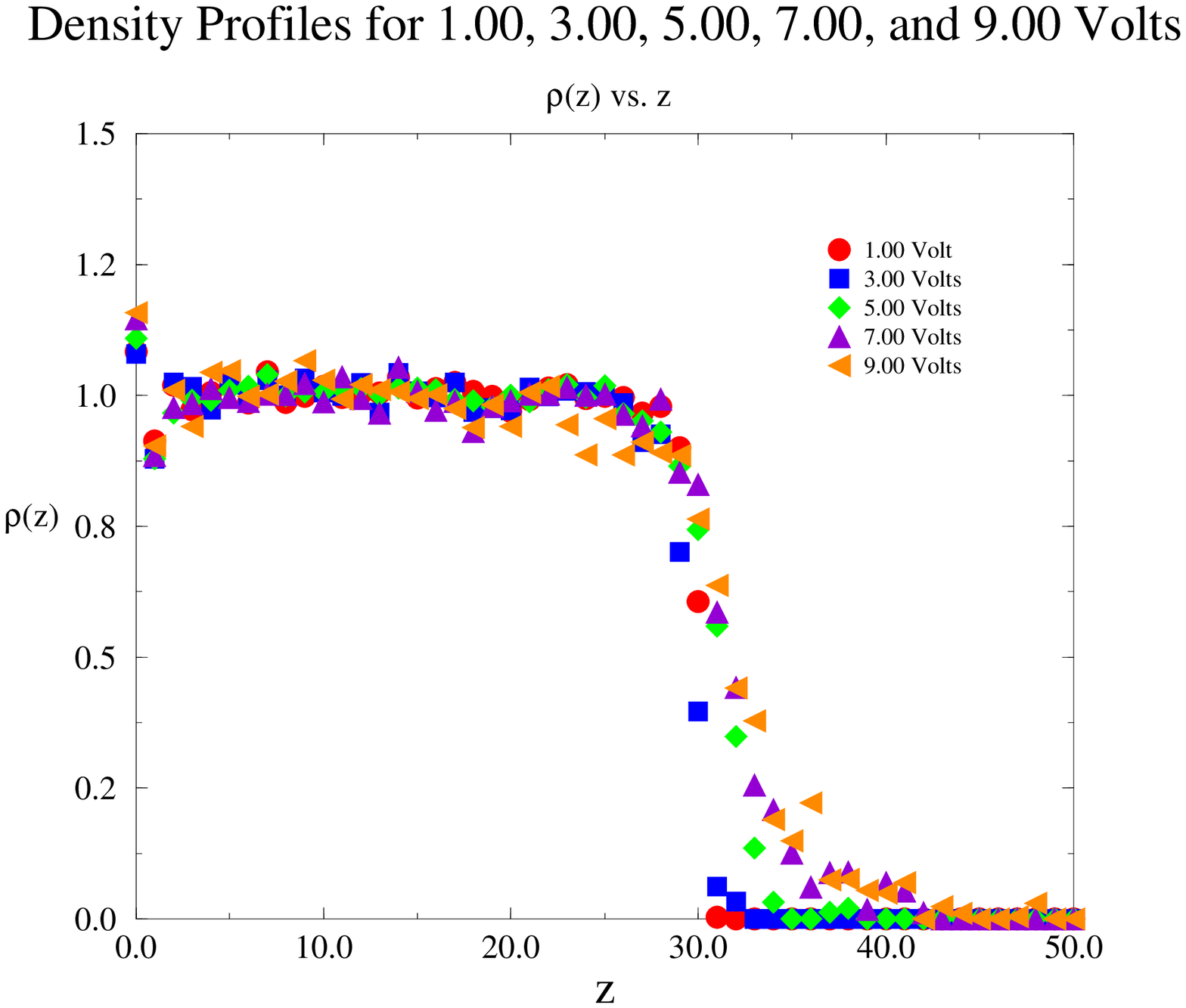}
}}

\newpage
\thispagestyle{empty}
\centerline{\hbox{
\psfig{figure=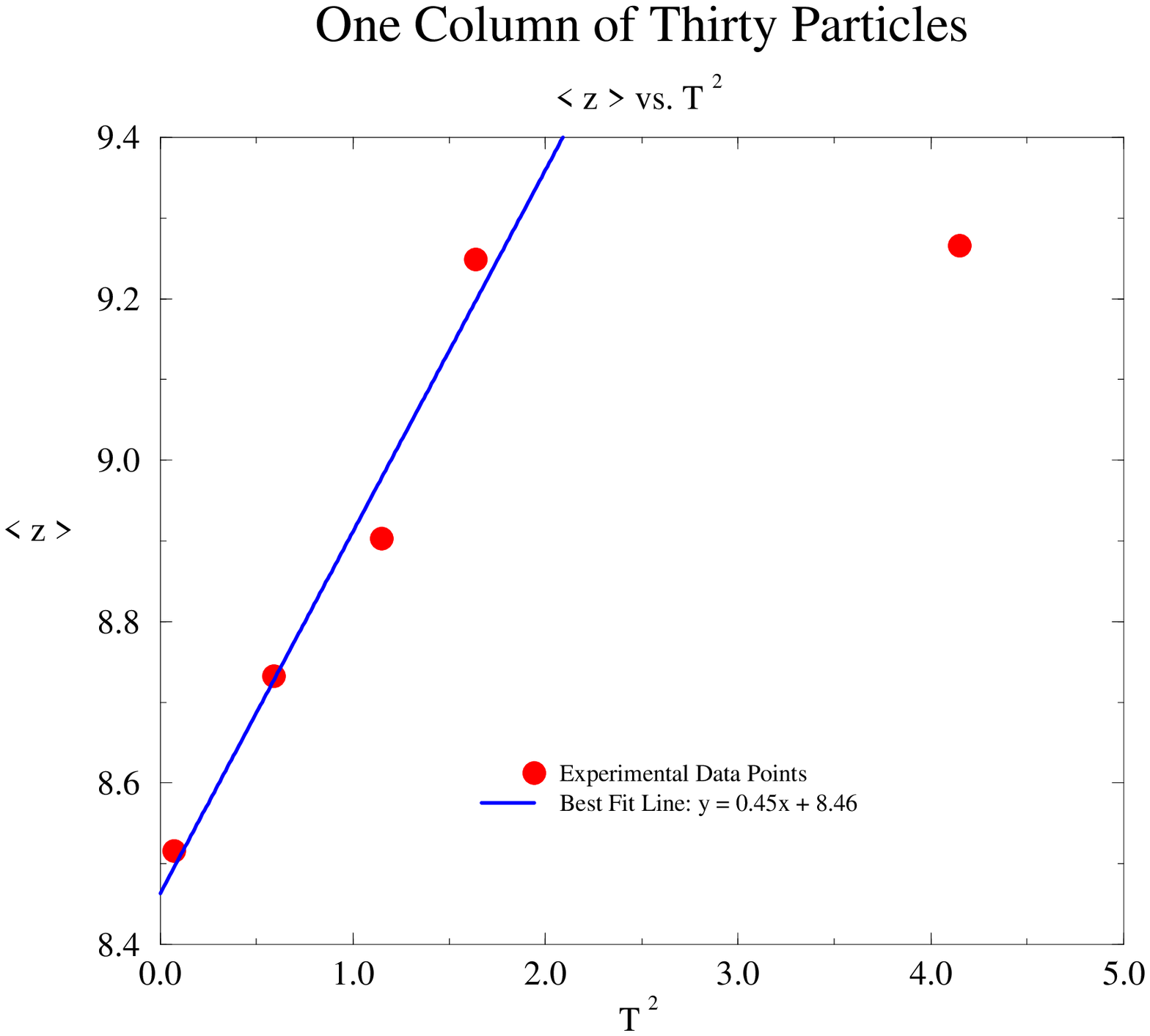}
}}
  
\newpage
\thispagestyle{empty}
\centerline{\hbox{
\psfig{figure=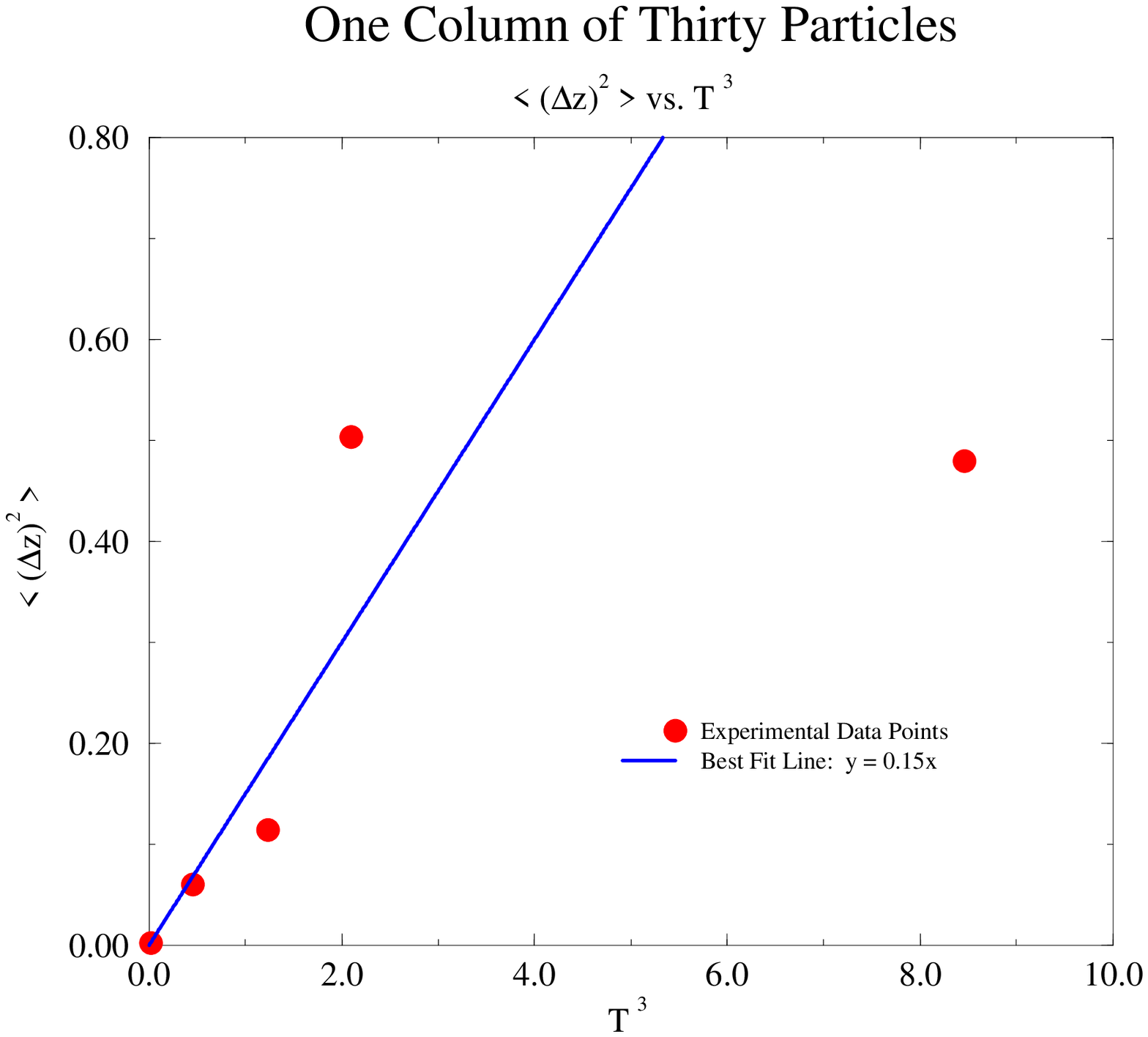}
}}

\newpage
\vskip 4.0 true cm  
\begin{tabular}{|c|c|c|} \hline

$V$ & $A$ & $\Gamma$ \\ \hline
\hspace{.3mm} 1.00 V \hspace{.3mm} & \hspace{.3mm} 0.045 cm \hspace{.3mm} 
& \hspace{.3mm} 2.90 \hspace{.3mm} \\ \hline 
3.00 V & 0.174 cm & 11.27 \\ \hline
5.00 V & 0.300 cm & 19.32 \\ \hline
7.00 V & 0.410 cm & 26.40 \\ \hline
9.00 V & 0.553 cm & 35.61 \\ \hline
\end{tabular}

\newpage
\vskip 4.0 true cm  

\begin{tabular}{|c|c|c|c|} \hline

$V$ & \multicolumn{2}{c|}{Parameters} & $R = \frac{\mu D}{\Gamma A}$ \\ \hline
    & $\mu$ & $T$&  \\ \hline
\hspace{.3mm} 1.00 V \hspace{.3mm} & \hspace{.3mm} 29.800 \hspace{.3mm} 
& \hspace{.3mm} 0.26800 \hspace{.3mm} & \hspace{.3mm}131.07   \hspace{.3mm}
\\ \hline 

3.00 V & 29.612 & 0.77000 & 8.62 \\ \hline
5.00 V & 31.182 & 1.07299 & 3.09 \\ \hline
7.00 V & 31.800 & 1.27999 & 1.69 \\ \hline
9.00 V & 32.000 & 2.03750 & 0.93 \\ \hline
\end{tabular}

\newpage
\vskip 4.0 true cm  

\begin{tabular}{|c|c|c|} \hline

$V$ & $T$ & $T_o$ \\ \hline
\hspace{.3mm} 1.00 V \hspace{.3mm} & \hspace{.3mm} 0.26800 \hspace{.3mm} 
& \hspace{.3mm} 0.26899 \hspace{.3mm} \\ \hline 
3.00 V & 0.6759 & 0.77865 \\ \hline
5.00 V & 1.0821 & 1.07246 \\ \hline
7.00 V & 1.4859 & 1.27794 \\ \hline
9.00 V & 2.0375 & 1.50445 \\ \hline
\end{tabular}

\end{document}